\documentclass[aps,prl,twocolumn,groupedaddress,showpacs]{revtex4-1}
\usepackage{amsfonts}
\usepackage{amsmath}
\usepackage{amssymb}
\usepackage{graphicx,bm,units}

\begin{document}

\title{Topologically Protected Extended States in Disordered 
Quantum Spin-Hall Systems without Time-Reversal Symmetry}
\author{Zhong Xu$^1$}
\author{L. Sheng$^1$}
\email{shengli@nju.edu.cn}
\author{D. Y. Xing$^1$}
\author{Emil Prodan$^2$}
\email{prodan@yu.edu}
\author{D. N. Sheng$^3$}
\affiliation{$^1$National Laboratory of Solid State Microstructures and
Department of Physics, Nanjing University, Nanjing 210093, China\\
$^2$ Department of Physics, Yeshiva University, New York, NY 10016, USA\\
$^3$ Department of Physics and Astronomy, California State
University, Northridge, California 91330, USA}
\date{\today }

\begin{abstract}
We demonstrate the existence of robust bulk extended states 
in the disordered Kane-Mele model with vertical and horizontal 
Zeeman fields, in the presence of a large Rashba coupling. 
The phase diagrams are mapped out 
by using level statistics analysis and computations of the 
localization length and spin-Chern numbers $C_\pm$. $C_\pm$ are 
protected by the finite energy and spin mobility gaps. The latter 
is shown to stay open for arbitrarily large vertical Zeeman fields, 
or for horizontal Zeeman fields below a critical strength or at moderate disorder. 
In such cases, a change of $C_\pm$ is necessarily accompanied by the closing of 
the mobility gap at the Fermi level. The numerical simulations reveal sharp changes in 
the quantized values of $C_\pm$ when crossing the regions of bulk extended states, 
indicating that the topological nature of the 
extended states is indeed linked to the 
spin-Chern numbers. For large horizontal Zeeman fields, 
the spin-gap closes at strong disorder prompting a change 
in the quantized spin-Chern numbers without 
a closing of the energy mobility gap.
\end{abstract}

\maketitle

{\it Introduction.} 
Topological insulators (TI) are materials characterized by robust 
properties against smooth deformations and disorder. Representative 
examples of TIs are the Chern insulators (CI)~\cite{HALDANE:1988rh}, 
quantum spin-Hall (QSH) insulators~\cite{Kane:2005np,Kane:2005zw,Bernevig:2006hl,Koenig:2007ko} 
and the strong topological insulators~\cite{Fu:2007ti,Hsieh:2008vm}. 
Samples of these materials display robust conducting states at the edges 
while being insulators in the bulk. The edge modes are connected to robust 
metallic states residing in the bulk and away from the Fermi 
level~\cite{Prodan:2009od,Prodan:2009lo}. The robustness of the edge 
modes and the bulk metallic states is a manifestation of a bulk 
topological invariant and, as such, there has been a sustained effort 
focusing on understanding the bulk states in disordered TIs~\cite{Sheng:2006na,Onoda:2007xo,Essin:2007ij,Obuse:2007qo,Obuse:2008ff,Yamakage2010xr,Prodan2010ew,GuoPRB2010fu,Prodan2011vy,ProdanJPhysA2011xk}.

In two-dimensional bulk systems, the extended quantum states can survive localization only in extraordinary circumstances. 
In CIs, the bulk extended states carry a Chern number and, as such, they cannot be destroyed until these numbers are ``annihilated'' through collision with other extended states. In QSH insulators with time-reversal symmetry (TRS), it is believed that a similar scenario happens, but this time the bulk extended states 
are associated with the ${\bm Z}_2$ invariant~\cite{Kane:2005zw}, 
which is protected by the insulating gap and TRS. Consequently, TRS is believed to be crucial for delocalization in the bulk QSH insulators. The lack of a definitive answer is due to the lack of a complete theory of the ${\bm Z}_2$ invariant for aperiodic systems (for recent progress see~\cite{Loring2010vy,HastingsAnnPhys2011gy}). For example, the behavior of the ${\bm Z}_2$ invariant is not understood when the insulating gap is filled with dense localized spectrum, and the numerical approaches based on the twisted boundary conditions (BC) were not able to probe the diffusive regime, being limited to small systems~\cite{Essin:2007ij,GuoPRB2010fu}. Nevertheless, analyses based on direct methods such as level statistics~\cite{ProdanJPhysA2011xk} and the transfer matrix approach~\cite{Onoda:2007xo,Yamakage2010xr} have established beyond doubt that the bulk extended states are present in representative models of disordered QSH insulators with TRS. 

An alternative approach is provided by the spin-Chern invariants $C_\pm$~\cite{Sheng:2006na,Prodan:2009oh}. While, for QSH insulators with TRS, $C_\pm$ and the ${\bm Z}_2$ invariants predict perfectly consistent phase diagrams~\cite{LiPRB2010uv,ShanNJ2010,ProdanJPhysA2011xk}, they give conflicting predictions when TRS is broken. An early study~\cite{Onoda:2007xo} on the bulk extended states in disordered QSH insulators concluded that the protection against localization comes from both topology and TRS: Lack of any of the two will result in the immediate destruction of the bulk extended states. This conclusion was aligned with the predictions based on the ${\bm Z}_2$ invariant  and was quickly accepted by the community. But recent studies on QSH-like phases with broken TRS revealed a more complex picture. Ref.~\cite{ZhouPRL2008vy} showed that robust gapless edge states can happen in QSH systems with TRS replaced by other symmetries. Yang $et$ $al.$~\cite{Yang2011ru} studied the Kane-Mele model~\cite{Kane:2005np,Kane:2005zw} with a Rashba coupling and vertical Zeeman field, 
breaking both TRS and inversion symmetry, and found a topological QSH-like phase characterized by spin-Chern numbers that remain quantized until the bulk energy gap closes. 
The quantization of $C_\pm$ indicates that the topological order of the QSH systems 
is intact, and thus extended bulk states should exist in the systems, even when the TRS is broken.

In this Letter, we establish the existence of such extended states 
in disordered QSH models with strongly broken TRS. The phase 
diagrams of the models are explored using the level statistics 
analysis~\cite{MehtaRandMat2002}, the localization length 
calculations~\cite{MacKinnonPRL1981vu,MacKinnonZPB1983er} and 
the spin-Chern invariants~\cite{Sheng:2006na,Prodan:2009oh}.  
We show that the Kane-Mele model~\cite{Kane:2005np,Kane:2005zw} with a large vertical Zeeman field displays a nontrivial phase diagram in the $(E_F,W)$ plane ($E_F$ = Fermi level and $W$ = disorder strength), with a topological phase characterized by quantized spin-Chern numbers, surrounded by CI phases. {\it Each phase is completely surrounded by lines of robust extended states.} The situation remains the same for horizontal Zeeman fields below a 
critical strength. 
Beyond the critical strength, the spin-gap closes at
strong disorder, so that the extended states 
disappear without a closing of
the energy mobility gap. 
Our results can be consistently explained based on the 
non-commutative theory of the spin-Chern numbers.
 
 \begin{figure}
    \centering
    \includegraphics[clip,width=8cm, angle=0]{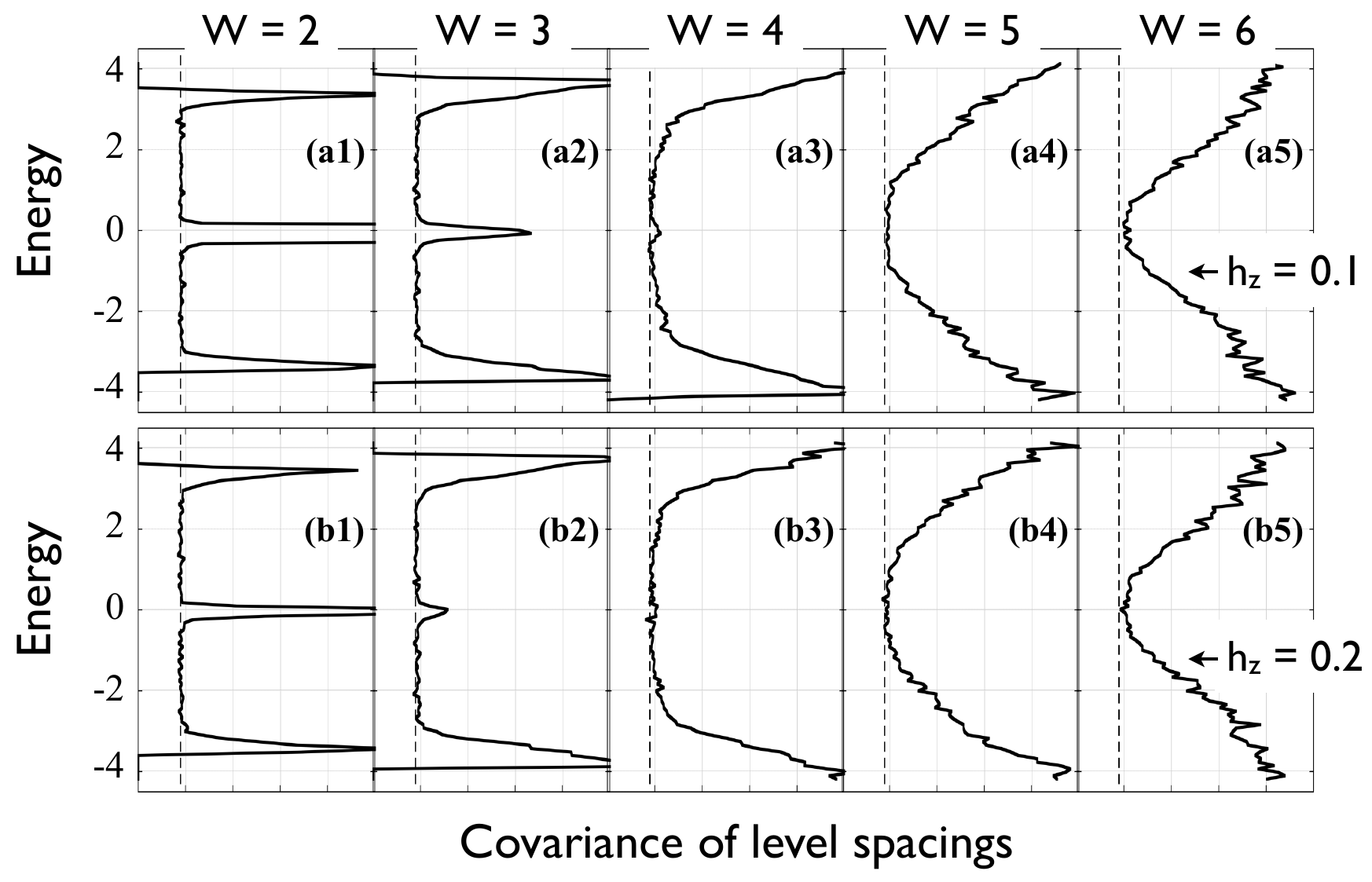}
\caption{Covariance of the level spacings ensembles collected at various $E$'s, $W$'s and for ${\bm h}=h_z\mbox{e}_z$ with $h_z=0.1$ and 0.2. The simulation was completed on a 40$\times$40 lattice and with 200 disorder configurations. The dotted line marks the 0.178 covariance of GUE. The windows span horizontally between 0 and 1.}
\label{Fig1}
\end{figure}

{\it The non-commutative spin-Chern Numbers.} In a lattice model with many-orbitals per site $|{\bm n},\alpha,\sigma\rangle$, where $\sigma$ ($=\pm 1$) and $\alpha$ represent the spin and other quantum numbers, let $\hat{\sigma}_z$ be the operator $\hat{\sigma}_z|{\bm n},\alpha,\sigma\rangle$=$\sigma|{\bm n},\alpha,\sigma\rangle$. If $P$ denotes the projector onto the occupied states of the Hamiltonian $H$, one can consider the operator $P \hat{\sigma}_z P$ and its spectral projectors $P_\pm$ onto the positive/negative spectrum. In the presence of disorder, one can associate non-commutative spin-Chern numbers~\cite{BELLISSARD:1994xj} to $P_\pm$~\cite{Prodan:2009oh}:
\begin{equation}\label{Chern}
C_\pm=2\pi i  \ {\mathbb E} \left \{ \mbox{tr}_0\big{ \{ }P_\pm \big{[} -i[\hat{x}_1,P_\pm],-i[\hat{x}_2,P_\pm] \big{]}\big{ \} } \right \},
\end{equation}
where ${\mathbb E} \{\ \}$ = disorder average, tr$_0$ = trace over the states at ${\bm n}$=${\bm 0}$ and $\hat{{\bm x}}$ = position operator. $C_\pm$ remain quantized and invariant as long as $\lambda_\pm^2={\mathbb E} \left \{ \mbox{tr}_0\big\{P_\pm {\bm x}^2 P_\pm \big \} \right \} <\infty$~\cite{Prodan:2010cz,ProdanJPhysA2011xk}. $\lambda_\pm$ can be viewed as localization lengths and $\lambda_\pm<\infty$  is enforced by a positive energy mobility gap and by the spin mobility gap in the spectrum of $P\hat{\sigma}_zP$. The localization length of $P$ will be denoted by $\lambda_{E_F}$.

\begin{figure}
 \includegraphics[clip,width=8.6cm, angle=0]{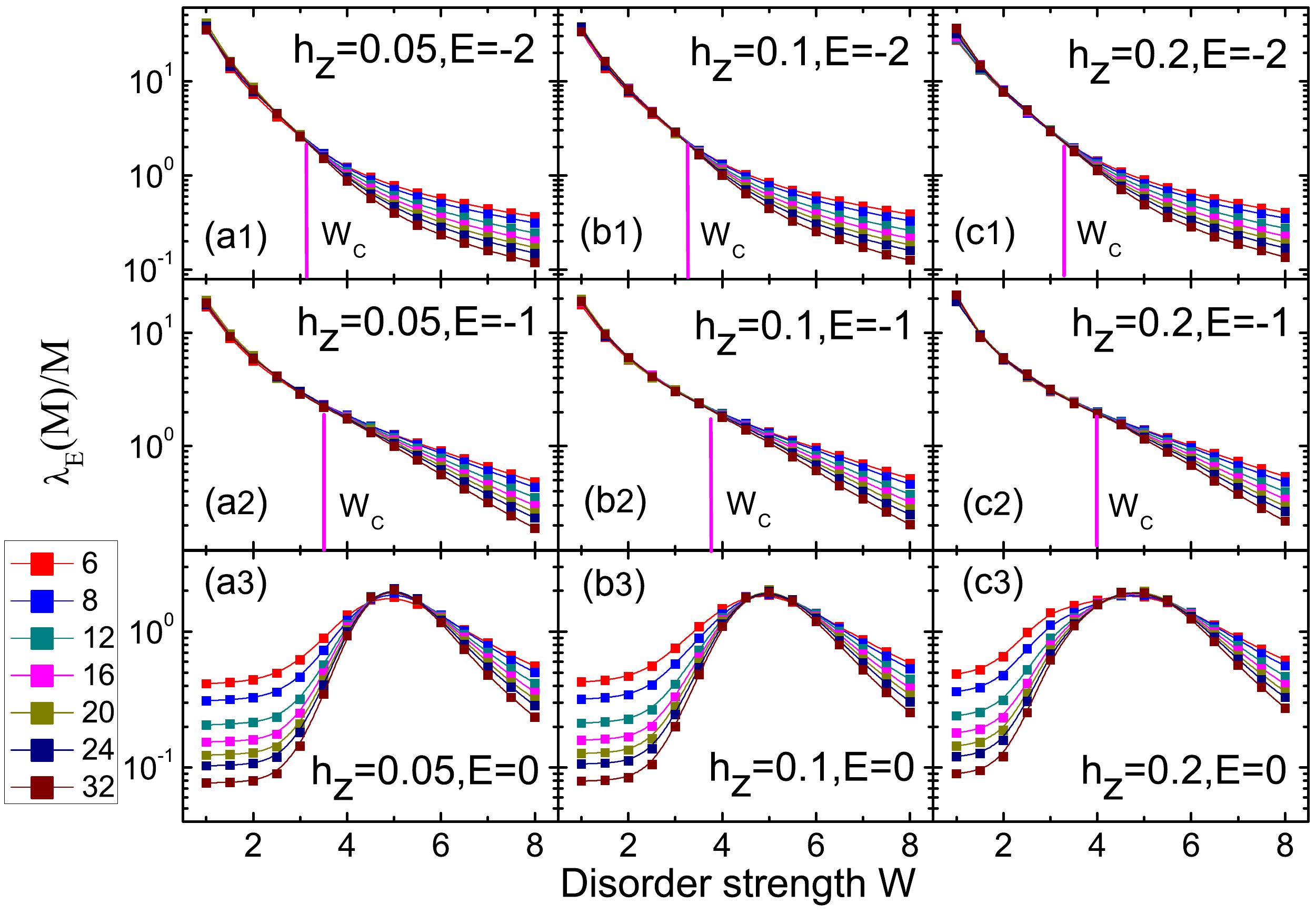}
\caption{Normalized localization length ${\lambda _E(M)}/M$ 
as a function of disorder strength $W$ at various energies $E$, calculated for a long tube
of $2M$-site circumference, in the presence of vertical Zeeman fields.  
Symbols with different colors correspond to different sizes. 
The system length is taken to be $10^6$ sites.}
\label{Fig2}
\end{figure}

Eq.~\ref{Chern} can be numerically evaluated using only one (periodic-) BC and one disorder configuration ($C_\pm$ are self-averaging)~\cite{ProdanJPhysA2011xk}. This and the precise quantization conditions make $C_\pm$ very effective tools~\cite{Prodan2011vy}. Note that in general $\frac{1}{2}\lambda_{E_F}^2\leq \lambda_-^2 + \lambda_+^2$, so $\lambda_\pm < \infty$ implies $\lambda_{E_F} < \infty$, and $\lambda_{E_F} \rightarrow \infty$ implies $\lambda_-$ or $\lambda_+ \rightarrow \infty$. Therefore, whenever $C_\pm$ are seen to take quantized values, one can safely conclude that the quantum states are localized at $E_F$. This allows one to map the regions of the phase diagrams that can harbor extended states, which are necessarily located in between the regions where $C_\pm$ assume different quantized values. 

{\it Analysis of the spin-gap.} Sharp predictions are possible when the behavior of the spin-gap is well understood. For clean QSH models with and without TRS, it is known that the spin-gap stays open when the system crosses the QSH phase boundary where the energy gap closes~\cite{ProdanJPhysA2011xk,LiPRB2010uv,Yang2011ru}. This remains true in the weak disorder regime and, as such, a sudden change of $C_\pm$ necessarily implies delocalization at $E_F$. For QSH models with TRS, the spin-gap is known to stay open at all disorder strengths~\cite{ProdanJPhysA2011xk}. Now, for systems with and without TRS, from $(P\hat{\sigma}_zP)^2=P\left (1-(i[\hat{\sigma}_z,P])^2 \right)P$
we see that the spectrum of $P\hat{\sigma}_zP$ near 0 is determined by the spectrum of $i[\hat{\sigma}_z,P]$ near $\pm 1$, the edges of its spectrum. Using $P=\oint_{\cal C} (\zeta-H)^{-1} \frac{d\zeta}{2\pi i}$, with ${\cal C}$ encircling the energy spectrum below $E_F$, one has
\begin{equation}\label{formula}
\begin{array}{c}
i[\hat{\sigma}_z , P]=\frac{1}{2\pi}\oint _{\cal C} (H-\zeta)^{-1}[\hat{\sigma}_z,H](H-\zeta)^{-1}d\zeta\ .
\end{array}
\end{equation}
Note that $[\hat{\sigma}_z,H]$ is independent of (non-magnetic-) disorder, so using the standard localization estimate ${\mathbb E}\{|\langle {\bm n}| (H-E)^{-1}|{\bm m}\rangle|^s\} \leq c_s e^{-s|{\bm n}-{\bm m}|/\lambda_E}$ ($s<1$, $c_s$ = generic constant, $\lambda_E$ = the localization length at $E$), which can be explicitly proven for many models~\cite{Aizenmann1993uf},  one can establish:
\begin{equation}
\begin{array}{c}
{\mathbb E} \left \{ |\langle {\bm n}| i[\hat{\sigma}_z,P]|{\bm m}\rangle | \right \} \leq c_s \overline{[\hat{\sigma}_z,H]}^{\frac{s}{2}} \lambda_{E_F}^2e^{-s|{\bm n}-{\bm m}|/4\lambda_{E_F}}\ ,\nonumber
\end{array}
\end{equation}
where $\bar{{\cal O}}$ = the largest matrix element of ${\cal O}$. Thus the averaged $i[\hat{\sigma}_z , P]$ decays exponentially and remains small when $[\hat{\sigma}_z,H]$ and $\lambda_{E_F}$ are both small (same for fluctuations). In such cases the spectrum of $i[\hat{\sigma}_z , P]$ is expected to be absent or localized at $\pm 1$. The conclusion is that, in general, whenever we see a sudden change in $C_\pm$ and $[\hat{\sigma}_z,H]$ is small, $E_F$ necessarily crosses 
 an energy region where $\lambda_{E_F}$ is large or infinite. 

Heuristically, we can go even further and make the following predictions: 1) The spin-gap is insignificantly affected by a vertical Zeeman term (in fact it is enhanced), thus, in a QSH system with only vertical Zeeman field, a sudden change in $C_\pm$ signals the divergence of $\lambda_{E_F}$; 2) For a horizontal Zeeman field, there is a direct effect on $i[\hat{\sigma}_z,P]$ and the spin-gap is expected to decrease and then close as the field is increased. Nevertheless, the spin-gap is expected to remain open for fields below a threshold value or at moderate disorder, where a sudden change in $C_\pm$ will signal again the divergence of $\lambda_{E_F}$.

\begin{figure}
 \includegraphics[clip,width=8cm, angle=0]{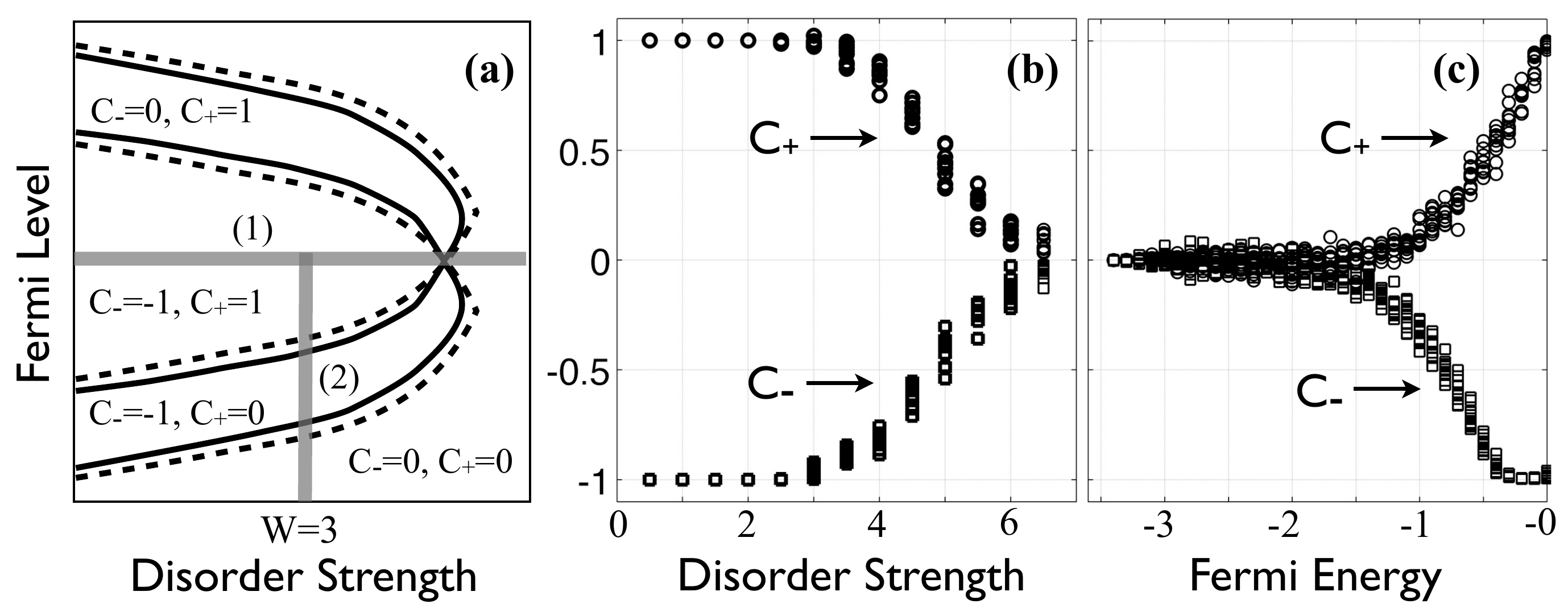}
\caption{(a) Phase diagram of a QSH insulator with a vertical Zeeman field. The phase boundaries (solid lines) harbor extended states. The dotted line shows the phase diagram when $V_R=0$. (b) and (c) show a computation of $C_\pm$ along the paths (1) and (2) (see panel a), respectively, completed on a 50$\times$50 lattice and with 10 disorder configurations.}
\label{Fig3}
\end{figure}

{\it Numerical results.} Our numerical simulations are based on the Kane-Mele model~\cite{Kane:2005np,Kane:2005zw}:
\begin{equation}\label{eq1}
\begin{array}{c}
 H =  \sum\limits_{ \langle nm \rangle} {c_n^\dag
{c_m}}+\frac{{2i}}{{\sqrt 3 }}{V_{SO}}\sum\limits_{ \langle\langle nm \rangle\rangle }
{c_n^\dag \hat{\bm \sigma}\cdot
({\bm d}_{km}\times {\bm d}_{nk}){c_m}} \smallskip \\
+i{V_R}\sum\limits_{\langle nm\rangle}{c_n^\dag\mbox{\bf e}_z\cdot
(\hat{\bm \sigma}
\times {\bm d}_{nm}){c_m}}
+\sum\limits_n c_n^\dag ({w_n} + {\bm h}\cdot\hat{\bm \sigma}){c_n}\ ,
\end{array}\nonumber
\end{equation}
where last term represents our addition of a spin independent, on-site potential $w_n$ with random amplitudes uniformly distributed in $\left[ { - W/2 , W/2} \right]$, and of a uniform Zeeman field that breaks the TRS. We 
will fix $V_{SO}$ and $V_R$ at 0.1, so that in the absence of the last term the model is inside the QSH phase.

Let us consider first a vertical Zeeman field ${\bm h}=h_z \mbox{\bf e}_z$. In Fig.~\ref{Fig1} we show the level statistics analysis for $h_z=0.1$ and 0.2 and various disorder strengths. The ensembles of energy level spacings were recorded from small energy windows centered at various energies $E$ and their covariance is plotted in Fig.~\ref{Fig1} as a function of $E$. For both $h_z$'s, one can see energy regions where the covariance is large, indicating localization, but also two regions where the covariance is practically equal to 0.178, the covariance of the Gaussian unitary ensembles (GUE), indicating the presence of extended states. Furthermore, the extended states are seen to drift towards each other and ultimately collide, annihilate and disappear as the disorder strength is increased. This 
is a strong indication that the extended states carry a topological number. 
For a system in the unitary class, one would expect extended states to 
occur at sharp discrete energies. This is precisely what was 
observed for CIs~\cite{Prodan2010ew,ProdanJPhysA2011xk}, but here 
we see in Fig.~\ref{Fig1} entire energy intervals where the covariance 
stays at 0.178, signaling a more complex phase diagram.  

\begin{figure}
    \centering
    \includegraphics[clip,width=8cm, angle=0]{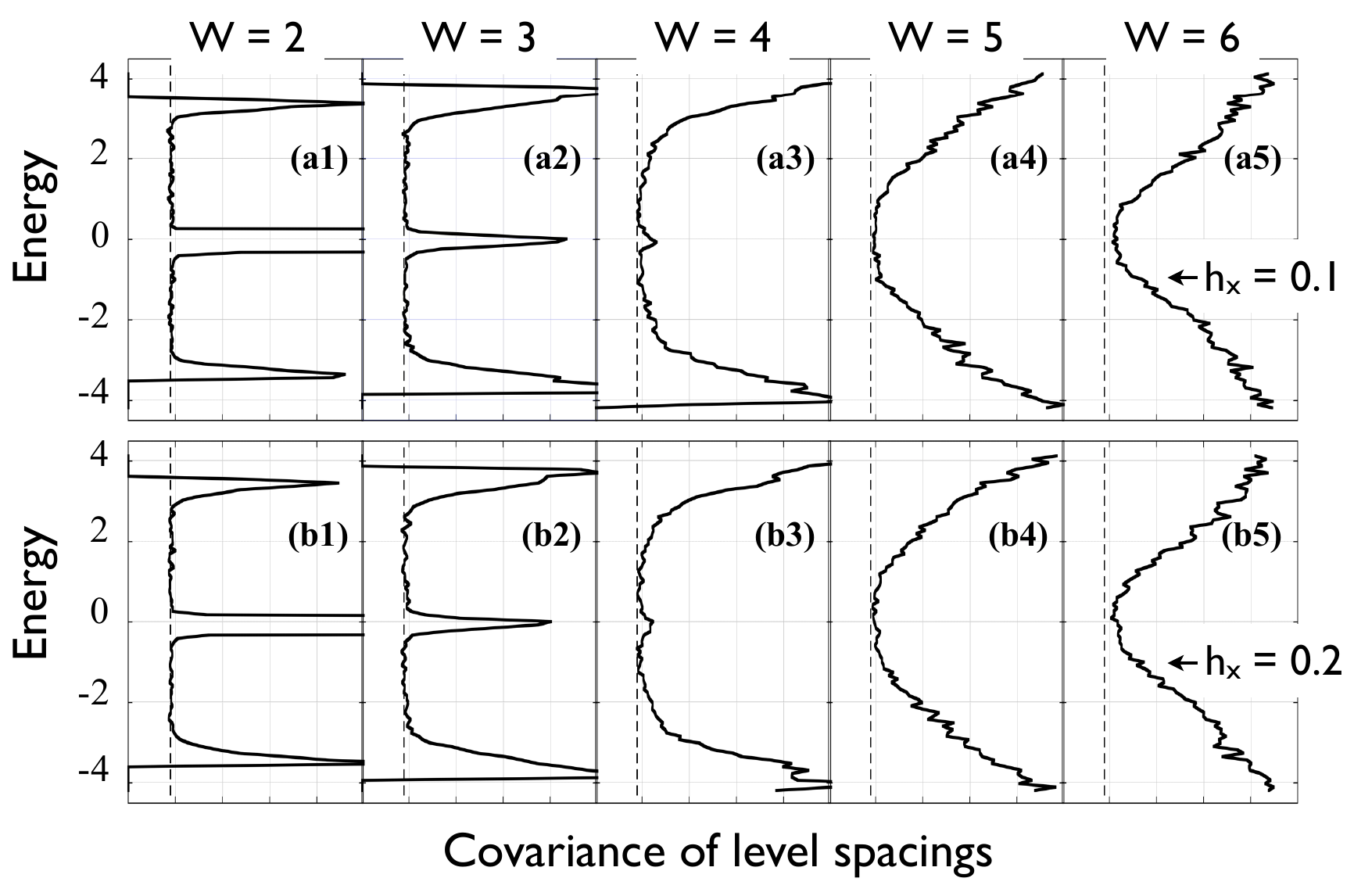}
\caption{Same as Fig.~1 but for horizontal Zeeman field.}
\label{Fig4}
\end{figure}

The delocalized character of the states can be firmly established by 
using the recursive Green's function approach~\cite{MacKinnonPRL1981vu,MacKinnonZPB1983er}. 
In Fig.~\ref{Fig2} we plot the calculated
normalized localization length ${\lambda_E(M)}/M$ as a function of $W$ 
for long tubes with circumferences ranging from $M=6$ to $32$, and for $h_z=0.05$, 0.1 and 0.2, 
and $E$=-2, -1 and 0. For each $E$, we can clearly see instances 
where ${\lambda _E(M)}/M$ decays sharply with $M$, but also where it remains nearly independent of $M$, essentially forming a line of fixed points. In the later case, $\lambda_E(M)$ grows linearly with $M$ therefore diverging in the thermodynamic limit. Consequently, the states are delocalized at these $E$'s and $W$'s. Furthermore, the independence of $\lambda_E(M)/M$ on $M$ indicates that we are dealing with critical states characterized by a vanishing $\beta$ scaling function~\cite{MacKinnonPRL1981vu,MacKinnonZPB1983er}, a characteristic of the Kosterlitz-Thouless type transition~\cite{KosterlitzJPC1973fr}. The extended states must have a topological origin, otherwise the system will belong to the trivial unitary class where all electron states are localized at positive $W$'s. The data in Figs.~\ref{Fig1} and \ref{Fig2} are consistent, for example, for $E=-2$ we see the covariance staying close to 0.178  even at $W=3$ and after that the extended states drift towards $E=0$. At about the same $W$, we see $\lambda_{E=0}(M)/M$ changing its behavior in Fig.~\ref{Fig2}(c1).  The extended states are seen to collide at about $W=5$ and $E=0$ in Fig.~\ref{Fig1}, and this is exactly where we see $\lambda_{E}(M)$ diverging in Fig.~\ref{Fig2}(a3-c3).

\begin{figure}
    \centering
    \includegraphics[clip,width=8.6cm, angle=0]{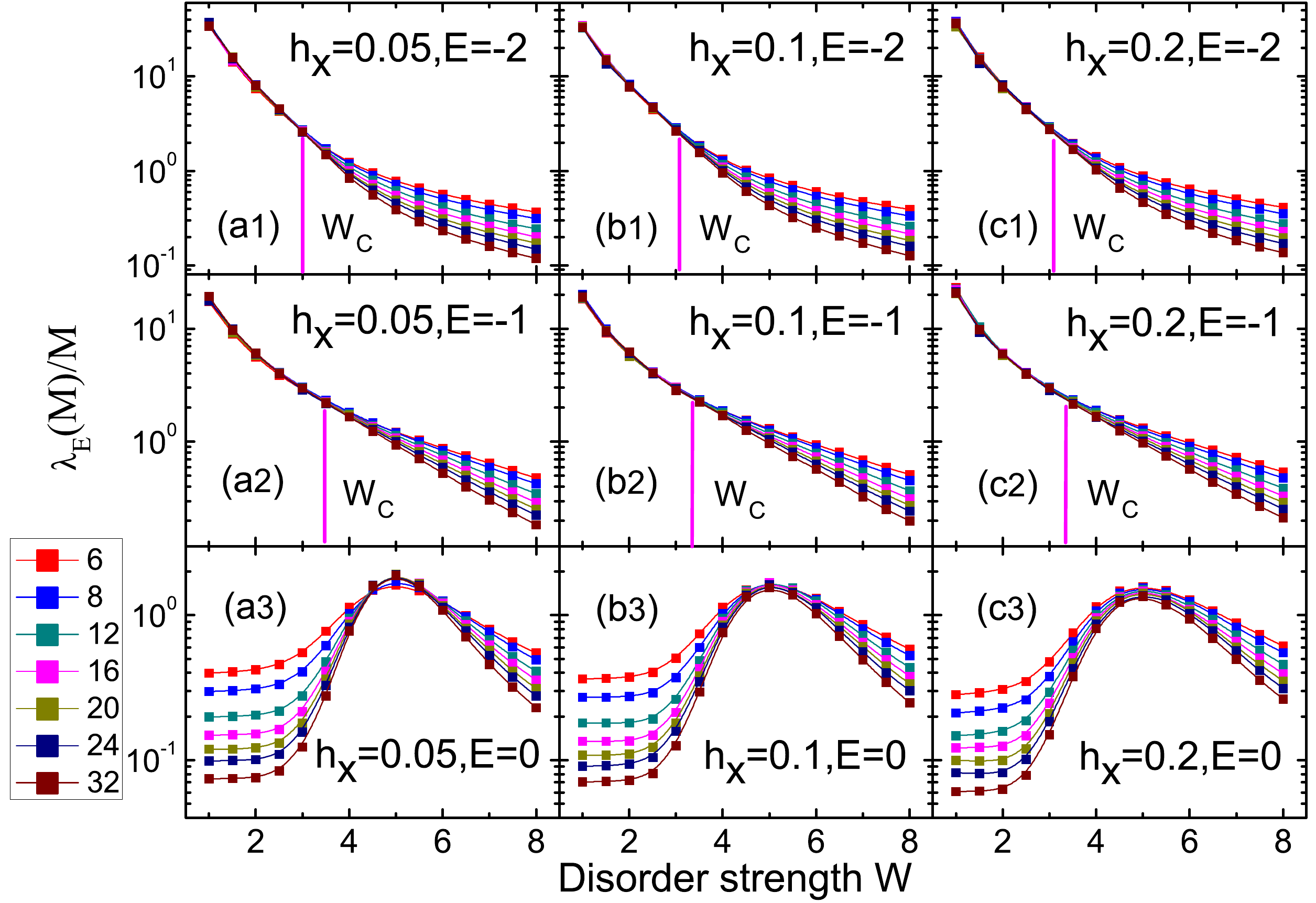}
\caption{Same as Fig.~\ref{Fig2} but for horizontal Zeeman field.}
\label{Fig5}
\end{figure}

To construct the phase diagram, it is convenient to start from the limit $V_{R}=h_z=0$, while holding $V_{SO}$ at 0.1. In this limit, the spin up/down sectors decouple and the phase digram in the $(E_F,W)$ plane, for each sector, consists of CI phases (with $C_\pm = \pm 1$) surrounded by a line of extended states and then by a trivial insulating phase~\cite{Prodan2010ew,ProdanJPhysA2011xk}. TRS assures that the phase diagrams for the spin up/down sectors overlap perfectly. As in Fig.~\ref{Fig3}(a), when $h_z$ is turned on, the phase diagrams shift revealing two regions where the total Chern number $C=C_- + C_+$ takes the values $\pm 1$, surrounding a region which can be characterized by $C_\pm=\pm 1$. The stability of the Chern numbers assures that the CI phases do not disappear when $V_R$ is slowly turned on but they could part away and open the phase diagram at large $W$'s. This cannot happen here because the spin-gap stays open so the $C_\pm=\pm 1$ and $C_\pm=0$ phases must be separated by extended states, a fact that can be seen explicitly in Fig.~\ref{Fig2}(a3-c3). We must conclude that the $C_\pm=\pm 1$ phase remains completely surrounded by the CI phases as illustrated in Fig.~\ref{Fig3}(a). The calculated values of $C_\pm$ in Fig.~\ref{Fig3}(b-c) for $h_z=0.2$, along the paths 1 and 2 in Fig.~\ref{Fig3}(a) further confirm our conclusions. Along path 1, both $C_\pm$ remain strictly quantized to $\pm 1$ until they simultaneously start an abrupt descent to 0. $C_\pm$ are seen to cross the values $\pm 0.5$ at about $W=5$, exactly where the annihilation is observed in Figs.~\ref{Fig1} and \ref{Fig2}. Along path 2, $C_-$ remains quantized at $-1$ well after $C_+$ started its descent to 0, indicating indeed the crossing of a region with total Chern number -1 (also confirmed by a direct calculation of $C$). This also provides direct evidence that the spin-gap remains open when crossing the boundary between $C_\pm=\pm 1$ and the CI phases, because its closing would have effected the quantization of both $C_\pm$ but in Fig.~\ref{Fig3}(c) $C_-$ clearly remains quantized. A last remark here is that, since the CI phases are surrounded by $C=0$ phases, they are necessarily surrounded by a line of extended states as illustrated in Fig.~\ref{Fig3}(a).

\begin{figure}
 \includegraphics[clip,width=8.6cm, angle=0]{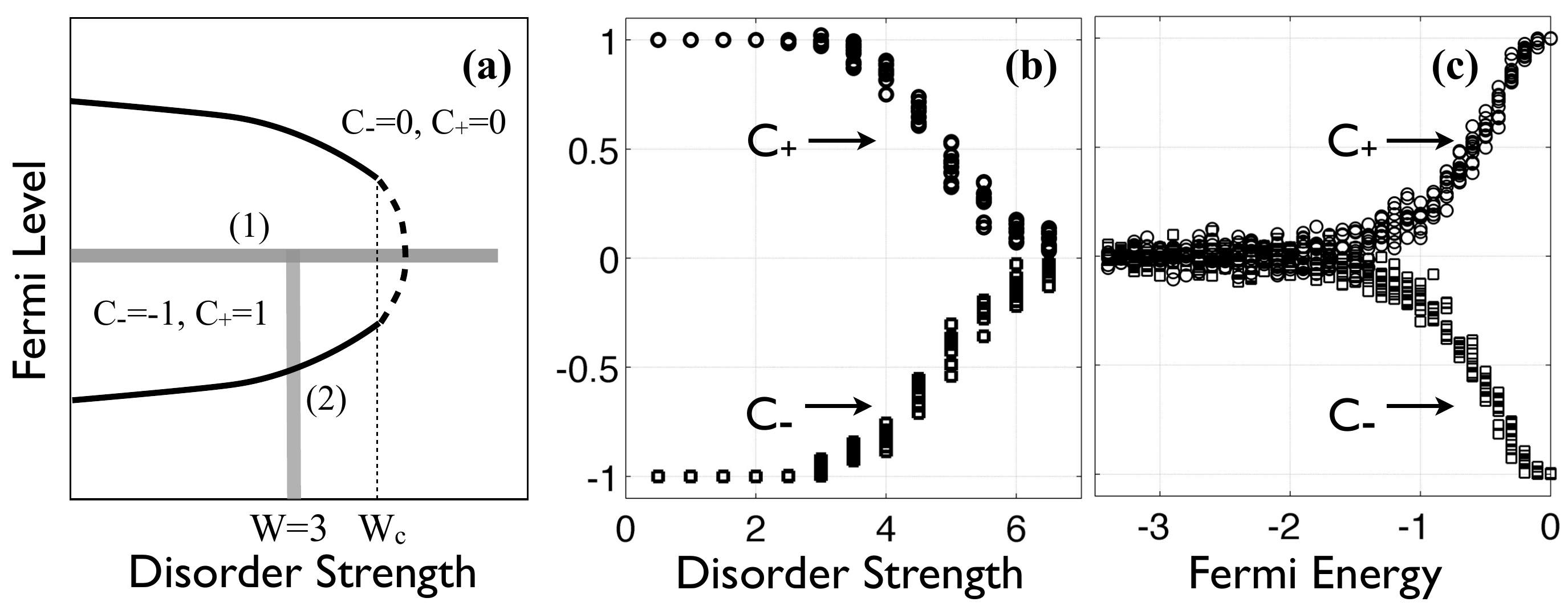}
\caption{Same as Fig.~\ref{Fig3} but for horizontal Zeeman field. The solid lines in panel (a) harbor extended states. The dotted line is where the spin-gap closes.}
\label{Fig6}
\end{figure}

We now consider the horizontal Zeeman field. In Fig.~\ref{Fig4} we show the level statistics analysis for ${\bm h}=h_x \mbox{\bf e}_x$ with $h_x=0.1$ and $0.2$. At smaller $W's$ we can again see regions where the covariance stays very close to 0.178, and we still see these regions drifting towards each other. However, the covariance is clearly seen to move away from 0.178 before the annihilation is complete (see panels a5 and b5). This is a first indication that the line of extended states is broken, which is indeed confirmed by the recursive Green's function analysis shown in Fig.~\ref{Fig5}. At $E=0$ for $h_x=0.1$ and 0.2, in panels (b3) and (c3) where the annihilation more or less takes place, we see $\lambda_E(M)/M$ decreasing with $M$, indicating a finite localization length. Nevertheless, both Figs.~\ref{Fig4} and \ref{Fig5} indicate that extended states are still present at lower $W$'s. This is also evident in Fig.~2 of Ref.~\cite{Onoda:2007xo}. Furthermore, the data in Fig.~\ref{Fig5} for the lower $h_x=0.05$ clearly demonstrate that the extended states survive here all the way till the annihilation point.

Based on the numerical data, on the additional computations of $C_\pm$ in Fig.~\ref{Fig6}(b-c) and on what we learned from the case with vertical field, we derived the phase diagram illustrated in Fig.~\ref{Fig6}(a) for large $h_x$ values. One can understand this diagram by starting from the diagram shown in Fig.~\ref{Fig3}(a) and rotating ${\bm h}$ from the $z$ to the $x$-direction. During this rotation, we expect the phase diagram to open at large $W$'s and the CI phases to separate and to continuously reduce size. Since there is an abrupt simultaneous change in the quantized $C_\pm$ along path 2, we have to conclude that all that is left when the rotation is completed is a doubly-degenerate line of extended states, which terminates abruptly at a critical $W_c$ (which is possible if the line is doubly degenerate). For $W$'s lower than $W_c$, the mobility spin-gap stays open and the change in the quantized $C_\pm$ when moving along path 2 must be due to the closing of the mobility energy-gap. From Figs.~\ref{Fig4} and \ref{Fig5}, we know that the mobility energy-gap stays open when moving along path 1, so the change in the quantized $C_\pm$ seen in Fig.~\ref{Fig6}(b) must be due to the closing of the mobility spin-gap. Therefore, as illustrated in Fig.~\ref{Fig6}(a), there must exist a line in the phase diagram where the mobility spin-gap closes, and this line together with the line of extended states completely encircle the region where $C_\pm=\pm 1$. Our final note here is that the change in the quantized $C_\pm$ brings measurable physical effects even beyond $W_c$, where we can still see energy regions with very large localization lengths, levitating and annihilating each other, in stark contrast with what happens in trivial insulators~\cite{Prodan2010ew,ProdanJPhysA2011xk}, where the extended spectrum simply fades away when the disorder is increased. 

{\it Summary.}
In conclusion, we used the spin-Chern numbers to predict the existence of robust bulk extended states in TIs 
with broken TRS and to map out the phase diagrams of the models. The predictions 
were well confirmed by the numerical computations of level statistics and localization lengths. Our study settles a long debated issue, namely, what protects the bulk extended states in TIs in the strong disorder regime? The answer is: topology alone. In contradistinction with the present common belief that the breaking of TRS in QSH insulators will lead to the sudden and complete localization of the bulk states, we found that QSH insulators with large Zeeman fields continue to display robust bulk extended states. We were able to derive the phase diagrams of the systems and, in the process, we discovered that the systems can be driven into topological CI phases by disorder and Zeeman fields.

\section*{Acknowledgments}

This work was supported by the State Key Program for Basic Researches of China under Grants Nos. 2009CB929504, 2007CB925104 (LS), 2011CB922103 and 2010CB923400 (DYX), and the National Natural Science Foundation of China under Grant Nos. 10874066, 11074110 (LS), and 11074109 (DYX). This research was also supported by the U.S. NSF grants DMS-1066045 and DMR-1056168 (EP), and by the U.S. 
DOE Office of Basic Energy Sciences under Grant No.  DE-FG02-06ER46305 (DNS).

\end{document}